\documentclass[amssymb,prb,aps,twocolumn,amsmath,showpacs]{revtex4}

\usepackage{graphicx}
\usepackage{amstext}
\usepackage{textcomp}
\usepackage{gensymb}

\begin{document}

\title{Spin-polarized triplet supercurrent in Josephson junctions with perpendicular ferromagnetic layers}
\author{Victor Aguilar}
\author{Demet Korucu} \altaffiliation[Permanent address: ]{Department of Physics, Gazi University, Ankara, Turkey}
\author{Joseph A. Glick} \altaffiliation[Present address: ]{IBM T.J. Watson Research Center, Yorktown Heights, NY 10598, USA}
\author{Reza Loloee}
\author{W. P. Pratt, Jr.}
\author{Norman O. Birge}
\email{birge@msu.edu} \affiliation{Department of Physics and
Astronomy, Michigan State University, East Lansing, Michigan
48824-2320, USA}

\date{\today}

\begin{abstract}

Josephson junctions containing three ferromagnetic layers with non-collinear magnetizations between adjacent layers carry spin-triplet supercurrent under certain conditions. The signature of the spin-triplet supercurrent is a relatively slow decay of the maximum supercurrent as a function of the thickness of the middle ferromagnetic layer.  In this work we focus on junctions where the middle magnetic layer is a [Co/Pd]$_N$ multilayer with perpendicular magnetic anisotropy (PMA), while the outer two layers have in-plane anisotropy.  We compare junctions where the middle PMA layer is or is not configured as a synthetic antiferromagnet (PMA-SAF).  We find that the supercurrent decays much more rapidly with increasing the number $N$ of [Co/Pd] bilayers in the PMA-SAF junctions compared to the PMA junctions.  Similar behavior is observed in junctions containing [Co/Ni]$_N$ PMA multilayers.  We model that behavior by assuming that each Co/Pd or Co/Ni interface acts as a partial spin filter, so that the spin-triplet supercurrent in the PMA junctions becomes more strongly spin-polarized as $N$ increases while the supercurrent in the PMA-SAF junctions is suppressed with increasing $N$.  We also address a question raised in a previous work regarding how much spin-singlet supercurrent is transmitted through our nominally spin-triplet junctions.  We do that by comparing spin-triplet junctions with similar junctions where the order of the magnetic layers has been shuffled.  The results of this work are expected to be helpful in designing spin-triplet Josephson junctions for use in cryogenic memory.
\end{abstract}

 \maketitle

\section{Introduction}

When a superconducting (S) metal is placed in contact with a non-superconducting or normal (N) metal, electron pair correlations extend into the normal metal, giving rise to the superconducting proximity effect \cite{deGennes1964}.  That proximity effect can extend into N over distances of several hundred nanometers at sufficiently low temperature, and indeed one can observe supercurrent in S/N/S Josephson junctions where the distance between the two S electrodes exceeds a micron \cite{Dubos2001}.  If the normal metal is replaced by a ferromagnetic material (F), then the situation changes drastically.  The pair correlations oscillate in sign as they penetrate into F, due to the momentum mismatch between the majority and minority spin Fermi surfaces in F \cite{Demler1997, Buzdin2005}.  Those oscillations decay algebraically in ballistic systems\cite{Buzdin1982}, and exponentially in diffusive systems \cite{Buzdin1991}.  The length scale governing the decay is typically very short -- less than one nanometer in strong F materials such as Co, Fe, or Permalloy (Py = Ni$_{80}$Fe$_{20}$) \cite{Robinson2006}.  Nevertheless, the oscillations give rise to interesting physics, including sign changes in the proximity-induced modification of the density of states \cite{Kontos2001} and in the sign of the current-phase relation in S/F/S Josephson junctions \cite{Ryazanov2001, Kontos2002}.

In 2001 Bergeret, Volkov, and Efetov predicted that it is possible to induce spin-triplet pair correlations in a conventional spin-singlet superconductor in contact with a ferromagnetic material with an inhomogeneous magnetization containing non-collinear elements \cite{Bergeret2001}.  Of the three spin-triplet components, those with z-component $m_s = \pm 1$ (often called "equal-spin triplets" in the literature) penetrate far into a homogeneous ferromagnet with its magnetization pointing along the z-axis.  The reason for this long-range proximity effect is that equal-spin triplets represent electron pairs where both electrons are in the same spin band -- either the majority band for $m_s = 1$ or the minority band for $m_s = -1$.  By now the predictions of Bergeret \textit{et al.} have been verified in a wide variety of different experiments; for reviews see \cite{Eschrig2011, Eschrig2015, Linder2015}.  The initial verification was achieved by measuring the supercurrent decay vs F-layer thickness in Josephson junctions; the signature of the induced spin-triplet pair correlations was that the supercurrent decayed much more slowly in junctions carrying spin-triplet supercurrent than in similar junctions that carried only the intrinsic spin-singlet supercurrent \cite{Keizer2006, Khaire2010, Robinson2010, Anwar2010}.

In this work we focus on Josephson junctions of the form proposed by Houzet and Buzdin \cite{Houzet2007}, which contain three magnetic layers with non-collinear magnetizations between adjacent layers.  In the years since the initial experimental results mentioned above, our group has focused on making Josephson junctions with increasing levels of control over the magnetic configurations inside the junctions.  In our initial work \cite{Khaire2010}, there was no magnetic control; spin-triplet supercurrent was generated haphazardly by the intrinsic non-collinearity existing in the as-grown junctions.  Shortly thereafter, we discovered that using an in-plane synthetic antiferromagnet (SAF) as the middle F layer \cite{Khasawneh2009} had an unforeseen advantage \cite{Klose2012}: after applying and then removing a large in-plane field to the junctions, the outer two layers became magnetized in the direction of the applied field, while the SAF underwent a spin-flop transition whereby the two magnetic layers making up the SAF ended up with their magnetizations perpendicular to the applied field direction.  That configuration with 90\degree angles between adjacent magnetizations optimizes the magnitude of the spin-triplet supercurrent.  In addition, having the outer two F layers with their magnetizations parallel means that the Josephson coupling across the junction has a $\pi$ phase shift everywhere, whereas the initial magnetic state with random domain orientations in the outer two layers produces a random pattern of 0 or $\pi$ Josephson couplings as a function of lateral position in the junction.  The combination of those two effects resulted in a factor of twenty enhancement of the spin-triplet supercurrent in the junctions compared to the as-grown state \cite{Klose2012}.  In another variation, we replaced the in-plane SAF with a [Co/Ni]$_n$ multilayer with perpendicular magnetic anisotropy (PMA) \cite{Gingrich2012}.  The PMA system has two advantages.  First, the magnetization of the PMA layer is naturally orthogonal to the magnetizations of the outer two layers, if the latter have the conventional in-plane magnetic anisotropy expected for thin magnetic films.  Second, the magnetic flux from the layer itself does not interfere with the ``Fraunhofer pattern" obtained when the supercurrent is measured as a function of magnetic field applied transverse to the current direction.  (The physics of magnetic fields applied to Josephson junctions will be discussed more in the next section.)  With the PMA system, we observed a very slow decay of the critical current in spin-triplet junctions, whereas the current decayed much faster in similar junctions carrying only spin-singlet supercurrent -- i.e. without the two outer magnetic layers.

Most recently, we replaced the PMA system with a PMA SAF, motivated by a desire to reduce stray fields produced by domain walls in the PMA system \cite{Glick2017}.  The spin-triplet junctions containing PMA SAFs have proven useful in demonstrating that the ground-state phase across the junction can be toggled between 0 and $\pi$ by reversing the magnetization direction of one of the outer two F layers \cite{Glick2018}.  However, the results from the junctions containing PMA SAFs raised two questions.  First, the decay of the critical current versus thickness of the PMA SAF was significantly steeper than we observed in similar spin-triplet junctions containing a simple PMA layer, but with a different materials system \cite{Gingrich2012}.  Second, at the smallest thicknesses of the PMA SAF, the critical current in the spin-triplet junctions was actually \textit{smaller} than in spin-singlet junctions containing the same PMA SAF, but without the two outer F layers that are necessary for the singlet-triplet conversion.  That raised the question: could we tell how much of the supercurrent in the spin-triplet junctions is actually carried by spin-triplet pairs rather than spin-singlet pairs?  Comparing the critical current in spin-singlet and spin-triplet junctions is not conclusive, because the latter junctions contain significantly more magnetic material than the former.

The purpose of this paper is to address both questions raised above.  We address the second question in section III, by comparing the critical current in our ``standard" spin-triplet junctions with the critical current in junctions containing the same magnetic layers, but with the order of the PMA SAF and one of the two in-plane ferromagnetic layers reversed. These so-called "shuffled" junctions are not capable of converting spin-singlet pairs to spin-triplet pairs and back again.  The spin-singlet supercurrent, on the other hand, is less sensitive to the order of the layers, so the critical current of such junctions provides a measure of the maximum spin-singlet supercurrent that can propagate through the spin-triplet junctions.  We observe that the supercurrent in the shuffled samples is significantly smaller than that in the standard junctions, confirming that the latter do indeed carry predominantly spin-triplet supercurrent.  We address the first question in section IV, by comparing the critical current in junctions containing the same PMA material system -- [Co/Pd]$_n$ multilayers -- either with or without the SAF configuration.  We observe that the critical current decays much less steeply with increasing $n$ in the junctions with the simple PMA system than in the junctions with the PMA SAFs.  We propose an explanation for this observation, namely that each Co/Pd interface acts as a partial spin filter for supercurrent.  As a result, the spin-triplet supercurrent becomes more strongly spin-polarized in the PMA system as the number $n$ of [Co/Pd] bilayers increases.  Polarization of the supercurrent hardly affects the magnitude of the supercurrent in the PMA junctions, while it causes a strong suppression of the supercurrent in the PMA-SAF junctions.

The questions addressed here are not just of academic interest.  Magnetic Josephson junctions with a controllable phase state have potential for use in cryogenic memory \cite{Dayton2018}.  While it is possible to design such junctions using only two F layers \cite{Golubov2002, Bell2004, Gingrich2016}, the constraints on the thicknesses of the F layers are more stringent in those junctions than in spin-triplet junctions \cite{Birge2019}.  Hence it is worthwhile to enhance our understanding of, and improve the performance of, spin-triplet junctions.  The junctions containing PMA SAFs, used in our demonstration of phase control \cite{Glick2018}, had very small critical currents -- approximately 5 $\mu$A for junctions with lateral area of 0.5 $\mu$m$^2$.  By replacing the PMA SAF with a simple PMA layer, we hope to achieve phase-controllable spin-triplet junctions with much larger critical currents.

\section{Junction Design, Fabrication, and Characterization}

Before continuing, we discuss the complications that arise when characterizing ferromagnetic Josephson junctions with an applied magnetic field.  If one applies a magnetic field to a standard Josephson junction (JJ) in a direction perpendicular to the current flow, the critical current $I_c$ oscillates in amplitude and decays as a function of field magnitude \cite{Barone1982}.  This is due to the contribution of the magnetic vector potential to the phase of the wavefunction, also known as the ``orbital effect."  In a rectangular junction, the behavior of $I_c(B)$ is similar to the well-known Fraunhofer pattern observed in single-slit optical diffraction, hence $I_c(B)$ data are often referred to that way even when the pattern is not a true Fraunhofer pattern.  (Note that the sin(x)/x function is squared in the optics context, whereas its absolute value describes $I_c(B)$ of a JJ.)  In circular junctions, the pattern is an Airy function.  Fraunhofer patterns acquired from S/I/S (I=insulator) and S/N/S junctions with appropriate geometry adhere closely to the theoretical ideal.  As soon as ferromagnetic materials are introduced into the junction, the situation gets more complicated.  The magnetic field and vector potential arising from the magnetization add to those associated with the external field.  In large-area junctions containing multiple domains of a strong F material, the resulting ``Fraunhofer" pattern is strongly distorted \cite{Bourgeois2001, Khaire2009}.  The problem is simplified in small junctions where the F layer has only a single magnetic domain; in that case the Fraunhofer pattern is shifted along the field axis by the nearly-uniform field produced by the F layer, so the central peak may be visible in a full field sweep.  But if the magnetization of the F layer switches direction before the central peak is reached, it may not be easy to extrapolate the data to the central peak to determine the true maximum value of $I_c$.  There are at least five ways to deal with this issue.  1) If one restricts oneself to weak ferromagnetic alloys with small domains, then the flux contributions from neighboring domains tend to cancel, and a reasonable Fraunhofer pattern emerges \cite{Ryazanov1999}.  Alternatively, if the F layer is very thin such that the magnetic flux in the junction due to the magnetization is much smaller than the flux quantum, $\Phi_0 = h/2e$, then the Fraunhofer patterns is hardly affected by the magnet.  2) One can replace a single F layer with a SAF to achieve flux cancellation domain-by-domain, resulting in a restored Fraunhofer pattern \cite{Khasawneh2009}.  3) One can fabricate extremely small JJs, so that the F layer tends to be single-domain and the Fraunhofer pattern is very wide in field; measurements at zero field then produce values of the critical current very close to the value at the maximum of the pattern \cite{Surgers2002, Bell2003, Robinson2006}.  4) In junctions with single-domain magnets, one can sometimes extrapolate the measured part of the pattern (before the magnetization switches direction) to estimate the height of the central peak in the pattern and thus the maximum critical current \cite{Baek2014, Niedzielski2015, Glick2017}.  5) If the magnetization points along the current direction, then the associated vector potential has little effect on the critical current, unless the junction is extremely long in the current direction \cite{Crosser2008,Scheer1997}.  The spin-triplet junctions described in this work benefit from two of the strategies listed above.  The PMA layer in the middle has its magnetization along the current direction; while the two outer magnetic layers are rather thin.  Nevertheless, we do observe shifts in the Fraunhofer patterns of our junctions due to the in-plane magnetizations of the outer two layers.

\begin{figure}[tbh!]
\begin{center}
\includegraphics[width=3.2in]{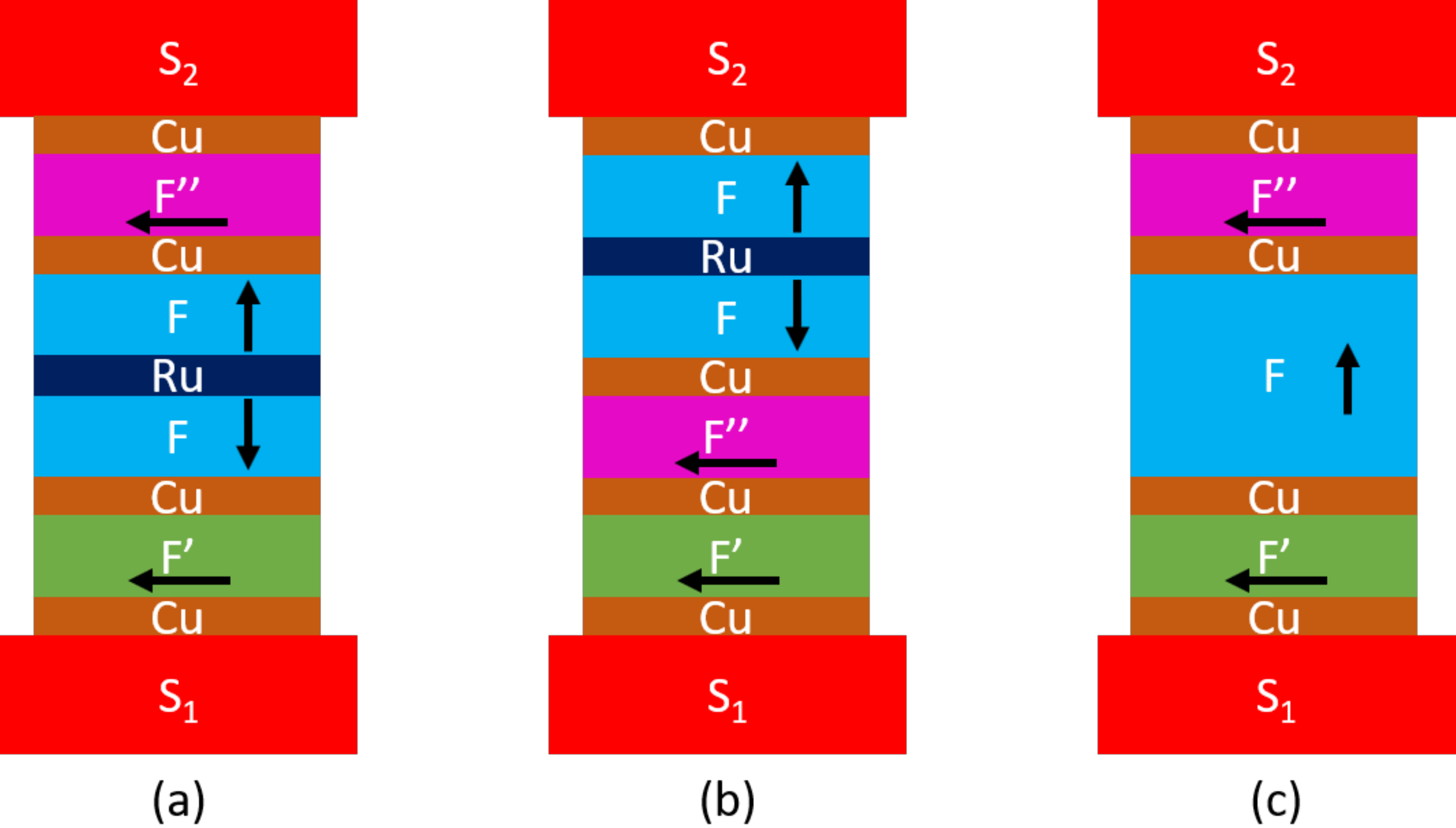}
\end{center}
\caption{Schematic diagrams of the three types of Josephson junctions used in this work (not to scale).  In all three cases, F is a [Pd(0.9)/Co(0.3)]$_n$ multilayer with perpendicular magnetic anisotropy (PMA), where all thicknesses are in nm and $n$ is the number of repeats of the Pd/Co bilayer. In (a) and (b), two Pd/Co multilayers are coupled antiferromagnetically by the Ru(0.95) spacer to form a synthetic antiferromagnet (SAF).  $F^\prime$ and $F^{\prime \prime}$ are Ni(1.6) with in-plane magnetization.  The superconducting base S1 is a multilayer of niobium and aluminum: [Nb(25)/Al(2.4)]$_3$/Nb(20) while S2 is Nb(150).  (a) Triplet PMA-SAF junction. (b) Shuffled PMA-SAF singlet junction.  (c) Triplet PMA junction.}
\label{TripletVsShuffled}
\end{figure}

A schematic drawing of our Josephson junctions with ``standard" spin-triplet JJ geometry, with the middle F layer a PMA SAF, is shown in Fig. 1(a).  The PMA nanomagnets labeled F in the figure are [Co(0.3)/Pd(0.9)]$_n$ multilayers (all thicknesses in nm), with Co adjacent to Ru and Pd adjacent to Cu.  The in-plane nanomagnets labeled $F^\prime$ and $F^{\prime \prime}$ in the figure are Ni(1.2) or Ni(1.6).  Ni is advantageous as a ``fixed" magnetic layer with in-plane magnetization, because it has high coercivity and higher transparency to Cooper pairs than most other magnetic materials used to date \cite{Baek2017}.  Fabrication of the junctions has been described in detail elsewhere \cite{Glick2017}; here we provide a brief summary.  First, we use photolithography to define the bottom lead pattern, and we sputter a nearly-complete stack consisting of the superconducting base layer and all of the magnetic layers.  The base is a Nb/Al multilayer of the form [Nb(25)/Al(2.4)]$_3$/Nb(20)/Cu(2), which we have determined to be smoother than a pure Nb base of the same thickness \cite{Wang2012}.  The magnetic heart of the PMA-SAF spin-triplet junctions is Ni(1.6)/Cu(4)/[Pd(0.9)/Co(0.3)]$_n$/Ru(0.95)/ [Co(0.3)/Pd(0.9)]$_n$/Cu(4)/Ni(1.6)/Cu(7)/Au(2).  The PMA junctions are nearly identical, except that the central Ru(0.95) layer is replaced by a Pd(0.9) layer so that the PMA system continues uninterrupted from one side to the other.  Because our sputtering system contains seven sputtering guns, we can sputter the entire stack for the PMA junctions without breaking vacuum. The PMA-SAF triplet junctions, however, contain eight different materials, hence we coat the base with a thin Au layer, break vacuum to replace the Nb target with Co, pump down again, and lightly ion mill the Au layer before depositing the magnetic stack. We then use electron-beam lithography with the negative e-beam resist ma-N2401 and Ar ion milling to define the junctions.  The junctions are elliptically shaped with lateral dimensions of 1.25 $\mu$m x 0.5 $\mu$m.  Ion milling is immediately followed by thermal evaporation of SiO$_x$ without breaking vacuum, to isolate the bottom and top superconducting electrodes.  After the e-beam resist if lifted off, top leads of Nb(150)/Au(15) are sputtered through another photolithographic stencil after gentle \textit{in-situ} ion milling of the protective Au layer. Cross-section TEM pictures of our junctions are shown in ref. \onlinecite{Glick2017}. Fabrication of the junctions shown schematically in Figures 1(b) and (c) proceeds similarly.

To characterize the junctions, we measure I-V curves as a function of field $H_{ext}$ applied parallel to the long axis of the elliptical junctions.  The I-V curves have the standard shape for overdamped junctions (RSJ model),\cite{Barone1982} except for some rounding for $I$ near $I_c$ visible in junctions with $I_c$ less than about 10 $\mu$A.  We fit the data using the standard form to extract $I_c$ at each value of $H_{ext}$.  To obtain $I_c(H_{ext})$, we first initialize the junctions with a field of $\mu_0 H_{ext}$ = -300 mT in the negative in-plane direction to magnetize the two Ni layers.  (Due to the very strong PMA of the [Co/Pd] multilayer system, a 300-mT in-plane field hardly perturbs the out-of-plane magnetization of the central PMA or PMA-SAF systems -- see Figure 1 of ref. \onlinecite{Glick2017}.)  After $H_{ext}$ is set to zero, the sample is slowly lifted above the liquid helium level in the measurement dewar, so that the Nb superconducting leads momentarily go into the normal state.  That process ensures that there is no magnetic flux trapped in the leads as a result of the initialization process.  After the junction is lowered back into the liquid helium, the field is stepped in the positive direction starting from -80 mT.  (There is no problem with trapped flux starting at this field value.)  Figure 2 shows a typical Fraunhofer pattern for one of the spin-triplet junctions containing a PMA-SAF.  The main central lobe and two side-lobes are clearly visible.  We typically stop the field sweep when the field passes through zero and reaches 20 mT in the direction opposite to the initialization direction. Continuing the field sweep past that point results in a distorted pattern due to the gradual switching of the two Ni layer magnetizations, as shown in Figure 6 of ref. \onlinecite{Glick2017}.  If the junction is initialized in the opposite direction, with $\mu_0 H_{ext}$ = +300 mT, then a down-sweep starting at +80 mT looks like the mirror image of the data shown in Figure 2.

\begin{figure}[tbh!]
\begin{center}
\includegraphics[width=3.2in]{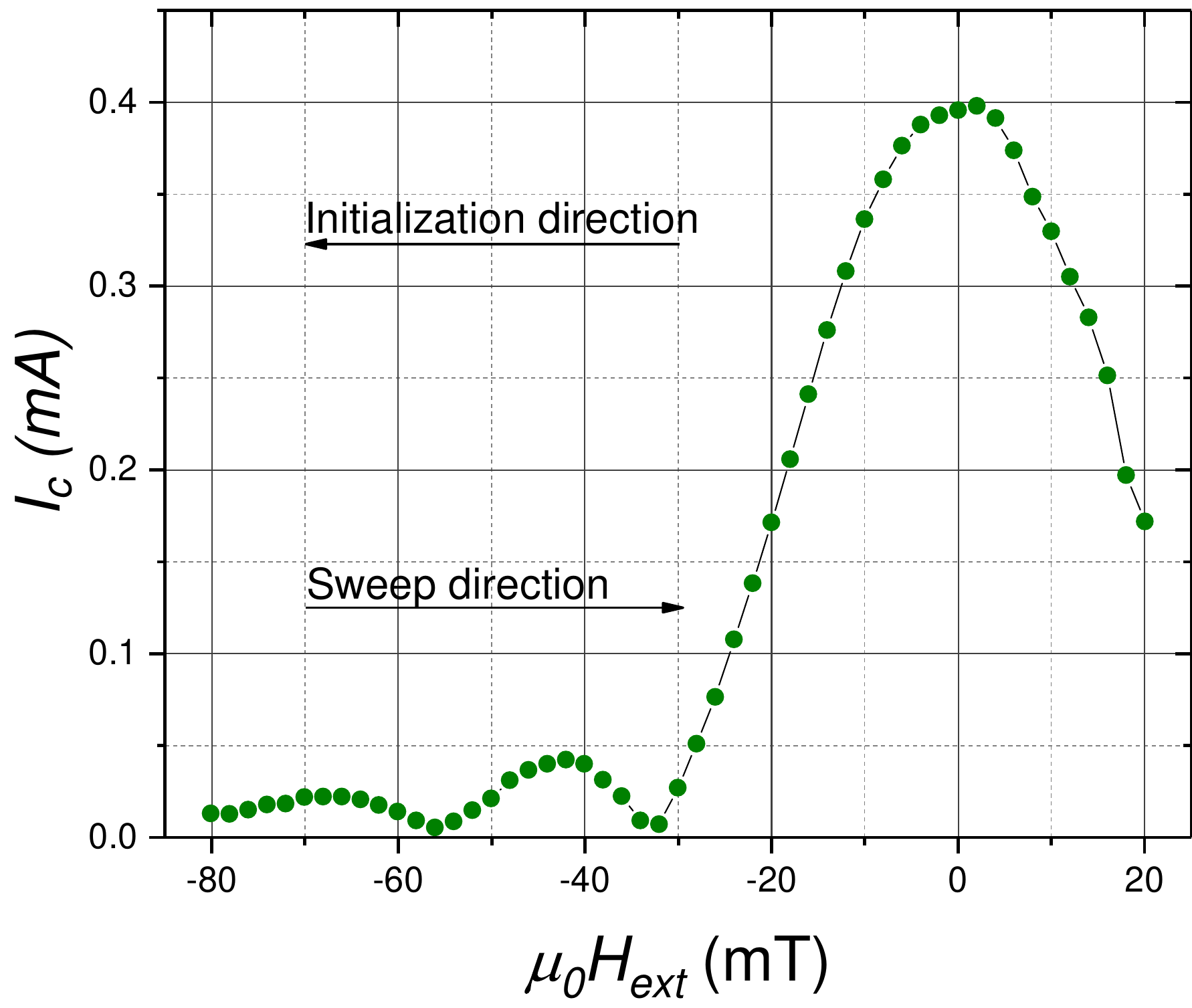}
\end{center}
\caption{Critical current versus applied magnetic field (``Fraunhofer pattern") for Josephson junctions containing three ferromagnetic layers.  Junctions are 1.25 $\mu$m x 0.5 $\mu$m ellipses with a 0.5 $\mu m^2$ area.  To initialize the Ni layers we magnetize the sample with a large (-300 mT) in-plane field at low temperature.  The sweep starts from the direction of initialization at -80 mT.  The field is increased in 2-mT steps; at each field we take an IV measurement in field.  The sweep is stopped at +20 mT before the magnetization of the Ni layer starts to reverse direction, which leads to distortions of the Fraunhofer pattern.}
\label{Fraunhofer}
\end{figure}

\section{Comparison of Spin-Triplet and ``Shuffled" Josephson junctions}

In a previous work,\cite{Glick2017} we compared the performance of PMA-SAF spin-triplet junctions with similar spin-singlet junctions -- the latter missing the $F^\prime$ and $F^{\prime \prime}$ layers needed to convert spin-singlet Cooper pairs from the superconducting electrodes into spin-triplet pairs in F.  Those data are reproduced in Figure 4 as the red triangles for the PMA-SAF spin-triplet junctions and the blue circles for the spin-singlet junctions. The plot shows the product of critical current times normal-state resistance, $I_cR_N$, vs the total number of [Pd/Co] bilayers, $N=2n$ for the PMA-SAF junctions.  As expected, the critical current in the spin-triplet junctions decays less steeply with $N$ in the spin-triplet junctions than in the spin-singlet junctions.  However, at the smallest value of $N=2$, the critical current is actually larger in the spin-singlet junctions than in the spin-triplet junctions.  That might seem surprising until one realizes that the spin-triplet junctions contain two additional ferromagnetic layers compared to the spin-singlet junctions.  Given the decline in supercurrent through each ferromagnetic layer as well as losses in transmission at interfaces due to Fermi-surface mismatch, it is no surprise that the spin-triplet junctions carry less supercurrent than spin-singlet junctions when $N=2$. At larger $N$, of course, the spin-triplet junctions win; at $N=6$ they already have over ten times as large a supercurrent as the spin-singlet junctions.  Given that spin-triplet junctions have the potential to be useful devices, however, it behooves us to know how much of the measured critical current in the junctions with $N=2$ is actually due to spin-singlet pairs.  To answer that question, we fabricated junctions of the form shown in Figure 1(b), which we call "shuffled" because the order of the ferromagnetic layers is shuffled compared to the spin-triplet junctions.  In these junctions the $F^\prime$ and $F^{\prime \prime}$ layers are adjacent to each other; hence they are not capable of providing spin-triplet current to the PMA layer and reconverting it to spin-singlet current after passing through the PMA layer. Since the shuffled junctions contain the same magnetic layers and the same number of interfaces as the spin-triplet junctions, they enable us to estimate of the amount of spin-singlet supercurrent that can pass through our PMA-SAF spin-triplet junctions.

Figure 3 shows a comparison of $I_cR_N$ for the spin-triplet and shuffled junctions, all with $N=2$.  The data for junctions with a PMA-SAF are on the right-hand side of the figure.  The data for the spin-triplet and shuffled junctions are represented by solid and open triangles, respectively.  Shuffled junctions were fabricated with two different Ni thicknesses, 1.2 and 1.6 nm.  As we anticipated, the values of $I_cR_N$ for the shuffled junctions lie well below those for the spin-triplet junctions. The data in the figure demonstrate that the amount of spin-singlet supercurrent that can propagate through this total combination of magnetic layers is indeed smaller than the spin-triplet supercurrent that can propagate through the same combination of layers when they are configured correctly to convert spin-singlet supercurrent to spin-triplet supercurrent and back again. The data on the left-hand side of Figure 3 are for similar junctions where the central F layer is a simple PMA layer with $N=2$ [Co/Pd] bilayers.  They are the topic of the next section.  For now, we point out simply that the values of $I_cR_N$ for the shuffled junctions (open hexagons) lie well below the values of $I_cR_N$ for the spin-triplet junctions (solid hexagons).  In addition to the data shown in Fig. 3, one PMA shuffled sample with $N=4$ was fabricated and measured.  Its critical current was barely measurable; our best estimate puts $I_cR_N$ at 0.003 $\mu$V.

\begin{figure}[tbh!]
\begin{center}
\includegraphics[width=2.5in]{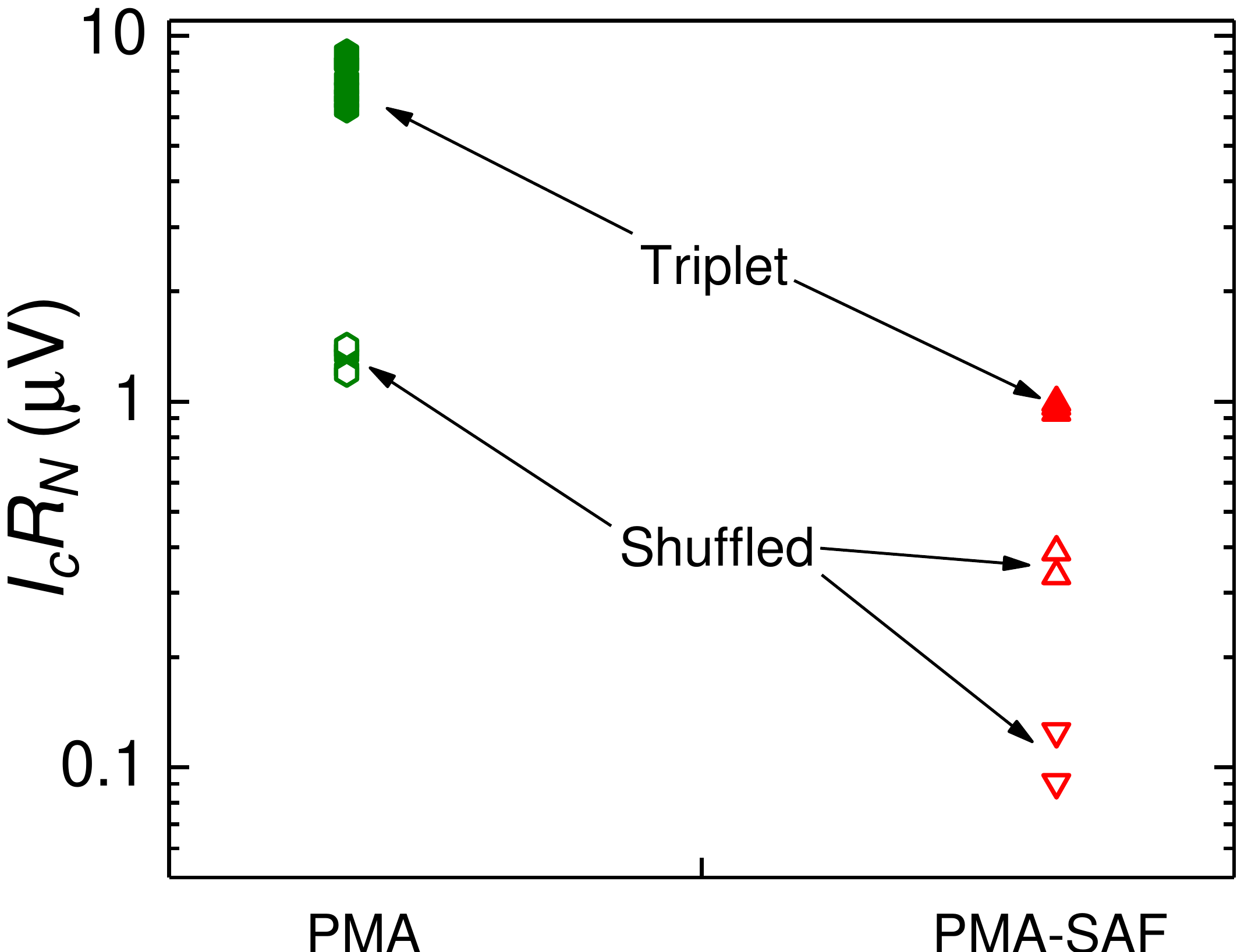}
\end{center}
\caption{(color online) Plot of $I_cR_N$ for spin-triplet junctions (solid symbols) in comparison with shuffled junctions (open symbols).  Left side: the central F layer is a simple [Pd/Co]$_n$ multilayer with PMA, as shown in Fig. 1(c).  Right side: the central F layer is a PMA SAF, as depicted in Fig. 1(a) and (b).  For the PMA-SAF shuffled junctions, two different Ni thicknesses were used: 1.2 nm (upward triangles) and 1.6 nm (downward triangles).  For all other cases the Ni thickness was 1.2 nm.}
\label{Shuffled}
\end{figure}

\section{Comparison of PMA-SAF and PMA spin-triplet Josephson junctions}

The primary goal of this work is to understand the difference in supercurrent transmission between spin-triplet junctions containing a PMA-SAF and those containing a single PMA layer of the same total thickness.  A schematic diagram of the PMA spin-triplet junctions is shown in Figure 1(c).  The measurement scheme was the same as that discussed previously for the PMA-SAF triplet junctions and the shuffled junctions.  The data for $I_cR_N$ as a function of the total number $N$ of [Co/Pd] bilayers in the PMA junctions are shown as green diamonds in Figure 4, while the data for the PMA-SAF junctions are shown as red triangles.  The data exhibit a striking difference between the PMA and PMA-SAF junctions.  The decay of $I_c$ with $N$ is much shallower in the PMA junctions than in the PMA-SAF junctions.  Least-squares fits of an exponential decay of the form $I_cR_N(N)=A_0 \textrm{exp}(-\alpha N)$ to the data in Fig. 4 give $\alpha_{CoPd}^{PMA} = 0.191 \pm 0.017$ for the PMA junctions, and $\alpha_{CoPd}^{PMA-SAF} = 0.751 \pm 0.047$ for the PMA-SAF junctions.  The ratio of those slopes is $3.9 \pm 0.4$.  In addition, the value of $I_c$ is larger in the PMA junctions than in the PMA-SAF junctions for all values of $N$, including the smallest value, $N=2$.  This result is very promising for practical memory devices, if we are able to fabricate controllable-phase PMA junctions by replacing one of the two Ni layers with NiFe, as we did with PMA-SAF junctions.\cite{Glick2018}

\begin{figure}[tbh!]
\begin{center}
\includegraphics[width=3.2in]{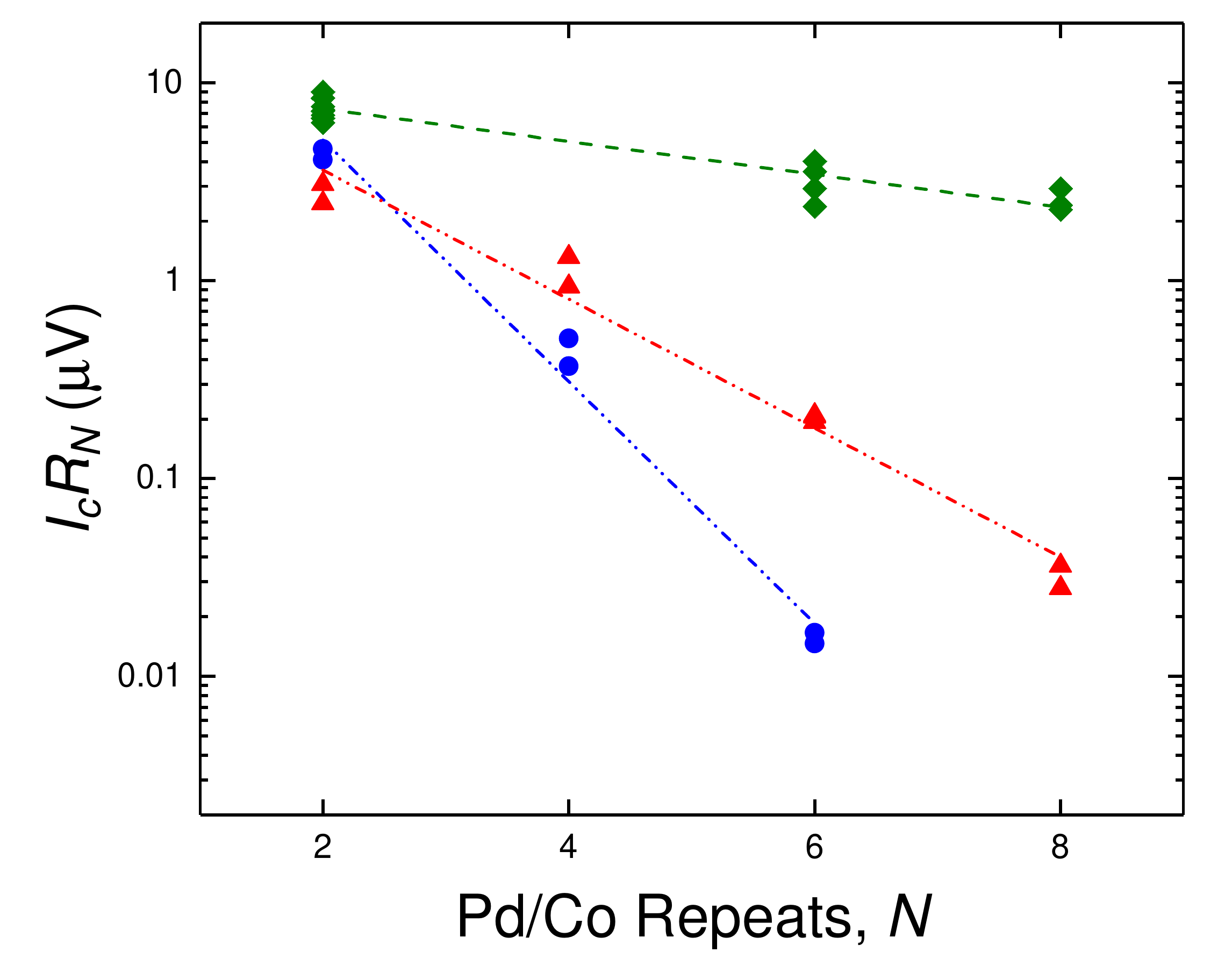}
\end{center}
\caption{(color online) Plot of $I_cR_N$ versus the \textit{total} number $N$ of Pd/Co repeats, i.e. $N=n$ for PMA junctions and $N=2n$ for PMA-SAF junctions.  Green diamonds: data from PMA triplet junctions, which have the largest critical current and shallowest decay with $N$.  Red triangles: data from PMA-SAF triplet junctions show a significantly steeper decay of $I_c$ with $N$.  Blue circles: data from PMA-SAF singlet junctions, which do not contain $F^\prime$ or $F^{\prime \prime}$ layers.  $I_c$ decays very rapidly with $N$ in the singlet junctions.  The data for the PMA-SAF and singlet junctions were published previously in Ref. [\onlinecite{Glick2017}].}
\label{IcRn}
\end{figure}

How to explain the striking results of Figs. 4?  Two theoretical works \cite{Volkov2010,Trifunovic2010} deal with precisely the kind of spin-triplet junctions studied here; the ``P-state" and ``AP-state" junctions in those works correspond to our PMA and PMA-SAF junctions, respectively.  (The fact that the F-layer magnetizations lie in the plane in the theoretical works, rather than out-of-plane as in our junctions, is not relevant; only the relative orientations of the different magnetizations matters.)  Both theoretical works predict that the critical current in the AP state will be slightly larger than in the P state, which clearly disagrees with the data shown in Fig. 4.  That is not surprising.  Both theoretical works utilize the quasi-classical theory, which does not account for differences in transport properties between majority and minority spin bands in the bulk of the F materials and at F/N interfaces.  The disagreement with experiment highlights the fact that the differences we observe between our PMA and PMA-SAF junctions are almost certainly due to those band structure effects.

Before we attempt to explain the results shown in Fig. 4, we quickly review the conceptual understanding of how three-layer spin-triplet junctions convert between spin-singlet a spin-triplet supercurrent \cite{Eschrig2011, Birge2018}.  Spin-singlet pairs entering $F^\prime$ from the bottom S electrode oscillate between the singlet and $m_s=0$ triplet states as they propagate through $F^\prime$.  When the pairs get to $F$ (and assuming that the $m_s=0$ triplet amplitude is not zero at that point), they are converted to $m_s=\pm 1$ states in the new, rotated basis determined by the magnetization direction of $F$.  If $F$ is sufficiently thick, then any singlet or $m_s=0$ triplet component entering $F$ will be completely suppressed after transmission through $F$.  When the $m_s=\pm 1$ pairs arrive at $F^{\prime \prime}$, they are converted back into the $m_s=0$ triplet state again by the basis rotation.  Finally, transmission through $F^{\prime \prime}$ converts the $m_s=0$ triplet state back into the singlet state, where it can be absorbed by the top S electrode.  The critical current $I_c$ of the junction is determined by how much of the singlet component remains after all of the basis transformations, Fermi-surface mismatches, and scattering that occurs both at the interfaces and in the bulk materials inside the junction.

Let us now focus on what happens inside the PMA system in the middle of the sample.  It is well-known in normal-state spintronics that a charge current passing through a ferromagnetic material F becomes partially spin-polarized due to the differences in density of states, Fermi velocity, and mean free path between majority and minority spin electrons.  It is undoubtedly true that F also acts as a spin filter with regard to the $m_s=+1$ and $m_s=-1$ spin-triplet electron pairs \cite{spinfilter,Moodera2007,Senapati2011,Bergeret2012}.  Let us assume that the majority band triplet component has higher transmission through a [Co/Pd]$_n$ multilayer than the minority band component.  The spin-triplet supercurrent passing through the multilayer then becomes partially spin-polarized, so that the $m_s=+1$ triplet component has larger amplitude than the $m_s=-1$ component.  In the PMA-SAF junctions, however, the majority-spin pairs in the first PMA multilayer become minority-spin pairs in the second PMA multilayer, and vice versa.  As a result, both triplet components suffer from the lower transmission amplitude of the minority component somewhere in the system.  If that were the whole story, then we would expect the $I_c$ data from the PMA-SAF samples to be shifted downward in Figure 4 with respect to the $I_c$ data from the PMA samples. But that is not what we observe. Rather, the slope of the data vs $N$ is steeper in the former, meaning that the downward shift increases with the number of bilayers $N$. That implies that the spin-filter efficiency of the PMA system increases with $N$, at least for $N$ as large as 8, which is the largest value probed in this work.

A possible explanation for the increase in spin-filter efficiency with $N$ is that each individual Co/Pd interface acts as a partial spin filter, due to a spin-scattering asymmetry at the Co/Pd interface favoring transmission of majority-spin pairs over minority-spin pairs.  Unfortunately, the spin-scattering asymmetry at the Co/Pd interface has not been measured.  Fortunately, however, the interfacial spin-scattering asymmetry has been determined for several other material interfaces by analyzing Giant Magnetoresistance data obtained from metallic multilayers and spin valves.\cite{Bass2011}  The analysis is based on the ``two-current series resistor (2CSR) model",\cite{ZhangLevy1991,LeePratt1993,ValetFert1993} which has been very successful in describing a large collection of experimental results.\cite{Bass2011,Bass2016}  The 2CSR model ascribes different effective resistivities to majority and minority-band electrons inside F materials, and different effective interface resistances at F/N or F$_1$/F$_2$ interfaces.  A good example is the Co/Ni interface, where the spin-scattering asymmetry was found to be quite large \cite{Nguyen2010}: the area-resistance product for majority-band electrons was found to be $AR_{Co/Ni}^{\uparrow} = 0.03 \pm 0.02 f\Omega m^2$ while the minority band interface resistance is much higher: $AR_{Co/Ni}^{\downarrow} = 1.00 \pm 0.07 f\Omega m^2$.  Hence the Co/Ni interface acts very strongly to spin-polarize a normal current, and we suspect it does the same for supercurrent.  In addition, the probability for an electron to flip its spin at the interface has also been measured.  It is expressed as P=1-exp$(-\delta)$, with $\delta_{Co/Ni} = 0.35 \pm 0.05$.

Fortuitously, we also have data for spin-triplet junctions containing [Co/Ni]$_n$ multilayers in the PMA and PMA-SAF configurations.  Those data are shown in Figure 5. The data show a similar trend as the data in Figure 4, namely that the decay of $I_cR_N$ with increasing number of Co/Ni bilayers for the junctions containing PMA SAFs is much steeper than the decay for junctions containing a simple PMA multilayer.  Fits of exponential decay to the data in Fig. 5 give $\alpha_{CoNi}^{PMA} = 0.063 \pm 0.011$ for the PMA junctions, and $\alpha_{CoNi}^{PMA-SAF} = 0.273 \pm 0.048$ for the PMA-SAF junctions.  The ratio of the slopes is $4.3 \pm 1.0$, similar to the ratio $3.9 \pm 0.4$ of slopes for the PMA and PMA-SAF data sets in Fig. 4.  An obvious difference between the [Co/Pd] results shown in Fig. 4 and the [Co/Ni] results shown in Fig. 5 is the huge vertical offset between the PMA and PMA-SAF data in Fig. 5.  The exponential fits in Fig. 4 have nearly identical intercepts for the PMA and PMA-SAF data, whereas the fits in Fig. 5 have intercepts that differ by a factor of about 30.  That large offset is due to several differences between the structures of the two kinds of junctions, which were fabricated several years apart under different circumstances.  First, the Co/Ni multilayers in the PMA junctions had thicknesses Co(0.2)/Ni(0.4), whereas those in the PMA-SAF junctions had thicknesses Co(0.3)/Ni(0.6).  Second, the [Co/Ni] multilayers in the PMA-SAF junctions had Co on the outside rather than Ni; from normal-state transport properties, we believe that Co/Cu interfaces are less transparent to supercurrent than Ni/Cu interfaces.  And third, the spacers separating the F layer from the $F^\prime$ and $F^{\prime \prime}$ layers were Cu(2)/Al(4)/Cu(2) trilayers rather than simple Cu(4) layers.  Those structural differences, while mostly responsible for the large vertical offset between the two data sets shown in Fig. 5, should not affect the slopes; hence we urge the reader to focus only on the slopes.

\begin{figure}[tbh!]
\begin{center}
\includegraphics[width=3.2in]{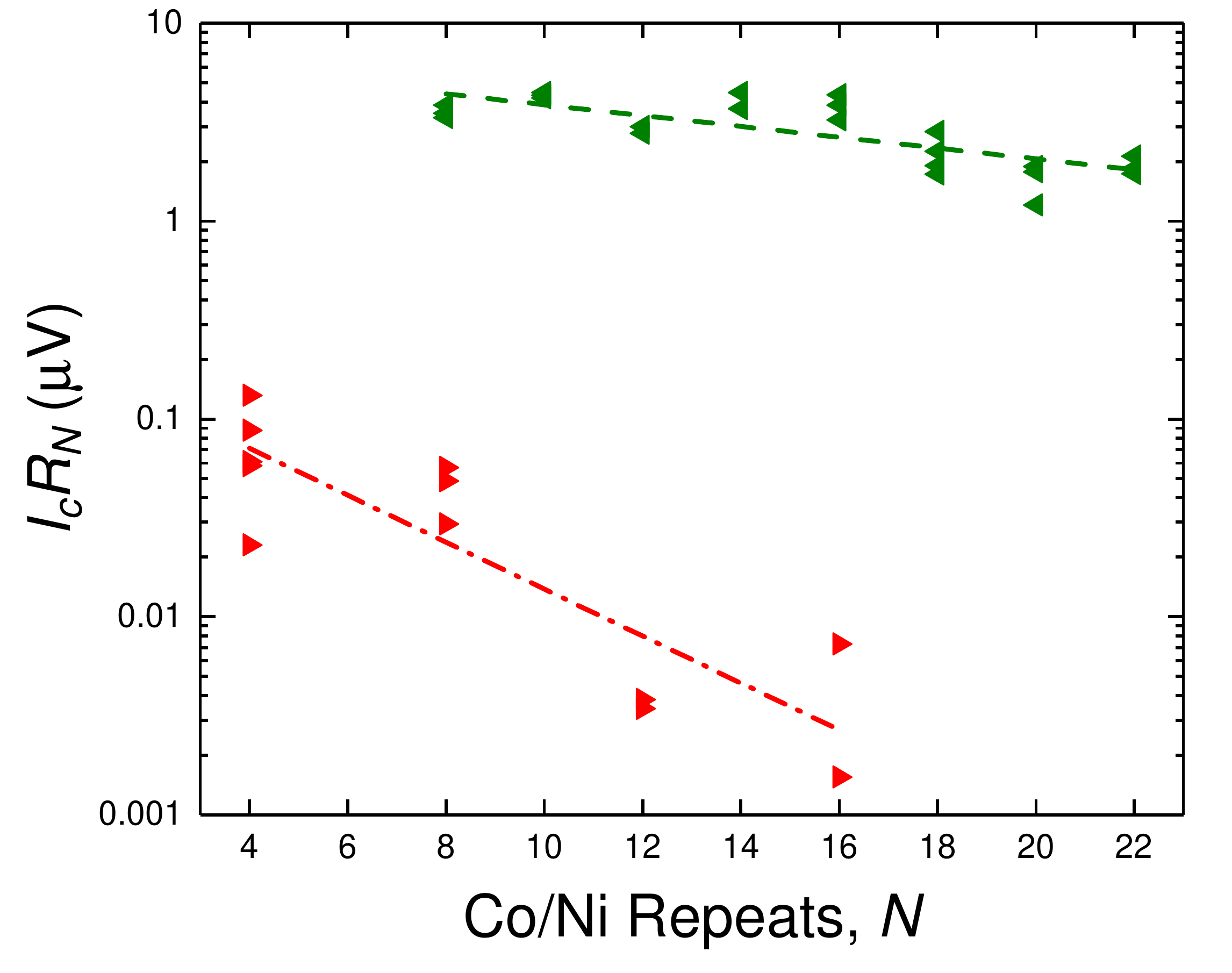}
\end{center}
\caption{(color online) Plot of $I_cR_N$ versus the \textit{total} number $N$ of Co/Ni repeats, i.e. $N=n$ for PMA junctions and $N=2n$ for PMA-SAF junctions.  Green left-pointing triangles: data from PMA triplet junctions taken from Ref. [\onlinecite{Gingrich2012}].  Red right-pointing triangles: data from PMA-SAF triplet junctions.  The large vertical offset between the two data sets is due mostly to several structural differences between the two kinds of junctions; hence we urge readers to focus on the different slopes of the two data sets.}
\label{IcRn}
\end{figure}

Is it possible to create a quantitative model to explain the slopes in Figs. 4 and 5 from the normal-state properties of the materials and interfaces in the junctions?  In principle, we believe that the answer is negative. While normal-state transport is dissipative and well described by the ``two-current series resistor model" mentioned above, supercurrent is a coherent process.  Nevertheless, we propose the following simple model as a guide to stimulate further theoretical work on this problem.

We start with the assumption that the total supercurrent through the junction is equal to the sum of the $m_s=+1$ and $m_s=-1$ triplet components of supercurrent in the central F layer of the junction.  (That assumption can be justified using the hand-waving model of supercurrent in refs. \cite{Eschrig2011,Birge2018}.)  We write that in shorthand as

\begin{equation}
I_c \propto T_{ms=+1} + T_{ms=-1}
\end{equation}

where $T_{ms=+1}$ and $T_{ms=-1}$ represent the transmission amplitudes of the $m_s=+1$ and $m_s=-1$ triplet components through the central F layer of the junction.  We will assume that each of those amplitudes can be expressed as a product of transmission amplitudes at the interfaces inside the F layer. (We neglect bulk effects for simplicity.)  Let $t_\uparrow$ and $t_\downarrow$ be the transmission amplitudes through a single Co/Pd interface for the majority-band and minority-band electron pairs, respectively.  A PMA junction with $N$ Co/Pd bilayers contains $2N$ Co/Pd interfaces, hence we have

\begin{equation}
T_{ms=+1}^{PMA} = t_\uparrow^{2N}
,\;
T_{ms=-1}^{PMA} = t_\downarrow^{2N}
\end{equation}

A PMA-SAF junction contains, in each half of the SAF, $N-1$ Co/Pd interfaces and one Co/Ru interface.  To keep the formulas as simple as possible, we will assume that the Co/Ru interface has the same properties as the Co/Pd interface, so we can say that each half of the SAF has $N$ Co/Pd interfaces.  (Deviations from that assumption will affect the intercept but not the slope of the line depicting $log(I_c)$ vs $N$.)  Since majority-band pairs in one half of the SAF become minority band pairs in the other half, we have:

\begin{equation}
T_{ms=+1}^{PMA-SAF} = T_{ms=-1}^{PMA-SAF} = t_\uparrow^{N}t_\downarrow^{N}
\end{equation}

Putting all this together gives us:

\begin{eqnarray}
I_c^{PMA} \propto t_\uparrow^{2N} + t_\downarrow^{2N}
\\
I_c^{PMA-SAF} \propto 2 t_\uparrow^{N}t_\downarrow^{N}
\end{eqnarray}

If we define the ratio $r = t_\downarrow/t_\uparrow$, then we can write

\begin{equation}
\frac{I_c^{PMA-SAF}}{I_c^{PMA}} = \frac{2r^N}{(1+r^{2N})}
\end{equation}

If $r^{2N} << 1$, then we get the simple result:

\begin{equation}
ln(I_c^{PMA-SAF}) - ln(I_c^{PMA}) = ln(2) + N ln(r)
\end{equation}

If each $I_c$ depends exponentially on $N$, as the data in both Figs. 4 and 5 show, then Eqn. (7) implies that the difference between the exponential slopes of the PMA and PMA-SAF junction data should be $ln(r)$. From the values for the Co/Pd interface, we found $-\alpha_{CoPd}^{PMA-SAF} +\alpha_{CoPd}^{PMA} = -0.751 + 0.191 = -0.56$, which gives $r_{CoPd}=0.57$.  For the Co/Ni interface, we found $-\alpha_{CoNi}^{PMA-SAF} +\alpha_{CoNi}^{PMA} = -0.273 + 0.063 = -0.21$, which gives $r_{CoNi}=0.81$.  (Note that, while the value of $r_{CoNi}$ is not much smaller than 1, $r_{CoNi}^{2N} = 0.19$ for the smallest value $N=4$ in Fig. 5, so our use of Eqn. (7) is justified.)  From the GMR data mentioned earlier, we would have expected $r_{CoNi}$ to be much smaller than 1.  One possible reason for that discrepancy is that the Co and Ni layers used in our junctions are too thin or too disordered to realize the bulk band structures of Co and Ni; hence the interface does not exhibit the large spin-scattering asymmetry determined from GMR data.  We note also that our model does not incorporate spin-flip and spin-orbit scattering, which suppress spin-triplet supercurrent, although we would expect them to cause equal suppression of $I_c$ in the PMA and PMA-SAF junctions.  Finally, we reiterate that the model presented here is undoubtedly too simple to describe quantitatively the supercurrent in a complex multilayered system.

\section{Conclusions}

In conclusion, we have measured the supercurrent in spin-triplet junctions containing ferromagnetic multilayers with perpendicular magnetic anisotropy (PMA).  We compared the behavior of the supercurrent in junctions where the PMA layer is or is not configured as a synthetic antiferromagnet (PMA-SAF).  The critical current decays much more steeply with increasing number of multilayer repeats $N$ in the PMA-SAF junctions than in the PMA junctions.  We attribute that difference to a strong increase in the degree of spin polarization of the supercurrent with increasing $N$, due to the cumulative spin-filtering effect of the interfaces within the PMA multilayers.  We also compared the supercurrent in spin-triplet junctions with that in junctions where the order of the layers was shuffled.  Because the shuffled junctions are not capable of converting spin-singlet supercurrent to spin-triplet supercurrent and back again, we believe that they provide an estimate of how much spin-singlet supercurrent is transmitted through the spin-triplet junctions containing the same number of kind of ferromagnetic layers.  As expected, the supercurrent is noticeably smaller in the shuffled junctions than in the triplet junctions even for $N=2$, and much smaller for $N=4$.  The results of this work are expected to be helpful in designing spin-triplet Josephson junctions for use in cryogenic memory.

Acknowledgements: We acknowledge fruitful conversations with F.S. Bergeret, A.Y. Herr, D.L. Miller, and N.D. Rizzo, and support from Northrop Grumman Corporation. We also thank B. Bi for technical assistance, and the use of the W.M. Keck Microfabrication Facility. This research was supported initially by the Office of the Director of National Intelligence (ODNI), Intelligence Advanced Research Projects Activity (IARPA), via U.S. Army Research Office contract W911NF-14-C-0115. The views and conclusions contained herein are those of the authors and should not be interpreted as necessarily representing the official policies or endorsements, either expressed or implied, of the ODNI, IARPA, or the U.S. Government.


\begin{thebibliography} {99}
\bibitem{spinfilter} The term ``spin filtering" often refers to tunneling through magnetic insulators -- either of single electrons [46] or of electron pairs in Josephson junctions [47,48].  But spin filtering is also an intrinsic property of metallic ferromagnets and their interfaces.
\end{thebibliography}
\end{document}